\documentclass[letter]{aa} 
\usepackage[colorlinks=true,citecolor=blue]{hyperref}
\usepackage{xcolor}
\hypersetup{
	colorlinks,
	linkcolor={red!50!black},
	citecolor={blue!50!black},
	urlcolor={blue!80!black}
}
\usepackage{multirow}
\usepackage{booktabs}
\usepackage{graphicx}
\usepackage{txfonts}
\begin{document}

    \title{Spectroscopy evidence for a so far unknown young stellar  cluster at the Galactic Center }
 \author{Á. Martínez-Arranz
	\inst{1}
	\and
	R. Sch\"odel
	\inst{1}
	\and
	F. Nogueras-Lara
	\inst{2}
	\and F. Najarro
	\inst{3}
	\and {R. Fedriani}
	\inst{1}
}
\institute{Instituto de Astrofísica de Andalucía (CSIC), University of Granada,
	Glorieta de la astronomía s/n, 18008 Granada, Spain\\
	\email{amartinez@iaa.es}
	\and
	European Southern Observatory, Karl-Schwarzschild-Strasse 2, 85748 Garching bei M\"unchen, Germany
	\and
	Centro de Astrobiología (CSIC/INTA), ctra. de Ajalvir km. 4, 28850 Torrejón de Ardoz, Madrid, Spain
}
\date{Received XXX; accepted YYY}
  \abstract
 { The nuclear stellar disk has been the most prolific star-forming region in the Milky Way over the last $\sim$30 million years. Despite hosting the massive Arches, Quintuplet, and Nuclear Stellar clusters, their combined mass below 10\% of the expected stellar mass formed in the region.This discrepancy, known as the "missing cluster problem," is attributed to factors such as high stellar density and tidal forces. Traces of dissolving clusters may exist as co-moving groups of stars, providing insights into the region's star formation history. Recently, a new cluster candidate associated to an HII region, was reported through the analysis of kinematic data}
 {Our aim is to determine if the young and massive stellar objects in the field share proper motion,  positions in the plane of the sky and line-of-sight distances, using reddening as a proxy for the distances.}
 {We reduced and analyzed integral field spectroscopy data from the KMOS instrument at the ESO VLT to locate possible massive, young stellar objects in the field. Then, we identified young massive stars with astrophotometric data from the two different catalogs to analyse their extinction and kinematics.}
 {We present here a group of young stellar objects, which share velocities, are close together in the plane of the sky and are locate at similar depth in the nuclear stellar disk}
 {The results presented here offer valuable insights into the missing clusters problem. They indicate that not all young massive stars in the Galactic center form in isolation; some of them seem to be the remnants of dissolved clusters or stellar associations.}
 
 \keywords{Galaxy: center, Galaxy: structure, Infrared: general, proper motions
 }
 \maketitle
%

\section{Introduction}
The Galactic center (GC), located at a distance of $\sim$\, 8.25\,kpc \citep{GC_distance_gravity}, harbors  the nuclear stellar disk (NSD), a flat rotating structure \citep{Sch2015, Ban_catalog} of about 150 pc across and 40 pc scale height \citep{Launhardt_2002,Sormani2022}.  \cite{three_cepheids} reported the presence of 3 classical Cepheids within the inner projected 40 pc of the GC. Their finding implies that about 10${^6}$ solar masses of stars were formed in this region over the past $\sim$10 Myr, a result that is also supported by the star formation history derived by \cite{paco_nature_2020}. With a star forming rate of $\in[0.2-0.8]$ M$_\odot$/yr in the last 30\,Myr \citep{three_cepheids, paco_nature_2020}, the NSD stands out for being the most prolific star forming region of the Galaxy when averaged over volume. This contrasts with the number of star clusters currently known in the NSD, the Arches and Quintuplet clusters, that add up to at most a few percent of this mass. This so-called missing clusters problem is not necessarily surprising because it can be explained by the extreme conditions present in the GC, that make the identification of young stars very challenging. The substantial and varying extinction within the GC \citep{extinction_los, paco_exctinction}, constrains observations to the near-infrared wavelength range. Moreover, the tidal field and tidal shocks can dissolve even the most massive cluster in $\lesssim$\,10\,Myr \citep{dissolve_GC,cluster_dissolution}, dropping its density below the stellar background, making the identification through overdensities impossible. This limitations makes it challenging to identify young clusters using color-magnitude diagrams (CMDs) which are dominated by reddening and provide little information on the intrinsic colors of stars, as highlighted in \cite{GNSI}. However, recent studies have discerned a substantial mass, several times 10$^{5}$ M$_{\odot}$, of young stars in Sgr B1 and Sgr C by examining luminosity functions \citep{Paco_B1, Paco_SgrC}. The presence of this extensive young stellar population within confined regions of the NSD supports the idea of coeval formation followed by rapid dissolution.

Identifying young stellar associations through spectroscopic studies is certainly a possibility. However, it is important to note that spectroscopy, especially at the required high angular resolution, typically involves a very limited field of view. Consequently, conducting comprehensive spectroscopic surveys across the entire NSD is impractical due to the significant time it would take to cover the entire region. Nevertheless, it is possible to narrow down our search for these stellar associations within the NSD to specific small areas by identifying co-moving groups \citep[see, for example,][]{Ban_cluster, comoving_arxiv}. To unveil the positions of previously undetected clusters or stellar associations, we can thus employ a two-step approach: first, identify co-moving groups, and then, confirm the presence of young massive stars and constrain the properties of these stars, and therefore of the cluster, by spectroscopy.
 
A co-moving group is characterized as a set of stars whose members are close in space and share similar velocities, exhibiting a velocity dispersion smaller than that of other stars in the field. It can be defined by a 6D parameter space, three dimensions for velocity components and three for position components. While the available catalogs for the GC provide position and proper motion data only in the plane of the sky, the third dimension in position, the line-of-sight distance, can be indirectly constrained by using observed stellar colors. Given the substantial extinction variability across the NSD \citep[][]{GNSIV, Ban_catalog, paco_NSD} and the almost constant intrinsic colors of observable stars, with variations mag $\lesssim$\,0.01 mag \citep[see Fig. 33 in][]{GNSI}, we propose that alterations in color primarily result from extinction \citep[see also][]{GNSIV}. Therefore, a group of stars exhibiting similar colors likely shares a comparable depth within the NSD \citep{paco_NSD}.

If a co-moving group is indeed the remnant of a cluster or a stellar association and is now in the process of dissolution, its member stars should still be relatively young, with ages on the order of $\lesssim$10 Myr. Massive, young stars (MYSs) can be separated from cool late-type stars via analysis of the spectral CO and Br$\gamma$ lines \citep{spec_class_1, spec_class_2}. Cool late-types present the $^{12}$CO(2,0), $^{12}$CO(3,1), and $^{12}$CO(4,2) bandhead absorptions at 2.30, 2.33, and 2.35 $\mu$m, respectively. Young, hot massive stars do not show these features and occasionally present an emission or absorption Br$\gamma$ line at 2.16 $\mu$m and/or  HeI abortion at  2.06 $\mu$m.

A co-moving group was reported in \cite{Ban_cluster}. This group was found in a known HII-emitting area \citep{HII_regions}, with the presence of a strong Paschen-$\alpha$ emitting star \citep{Blue_SG,Massive_stars}, making this region a solid candidate to host MYSs. Subsequently, we collected spectroscopy data of this regions using the Very Large Telescope (VLT) K-band Multi Object Spectrograph (KMOS) at the European Southern Observatory (ESO), under ESO program ID 105.20CN.001\footnote{\href{https://www.eso.org/sci/publications/announcements/sciann16012.html}{https://www.eso.org/sci/publications/announcements/sciann16012.html}}. Here, we analyzed the spectra of the objects in the field covered by the KMOS observations and selected the MYSs based on the  analysis of their spectra. Subsequently, we cross-matched these objects with two different catalogs: the GALACTICNUCLUES survey \citep{GNSI, GNSII} and the proper motion catalog by \cite{LIBRALA2021}, to obtain photometry and proper motions.

\section{Data and methods}

\subsection{Spectroscopy data}
\begin{figure}
	\centering
	\includegraphics[width=1\linewidth]{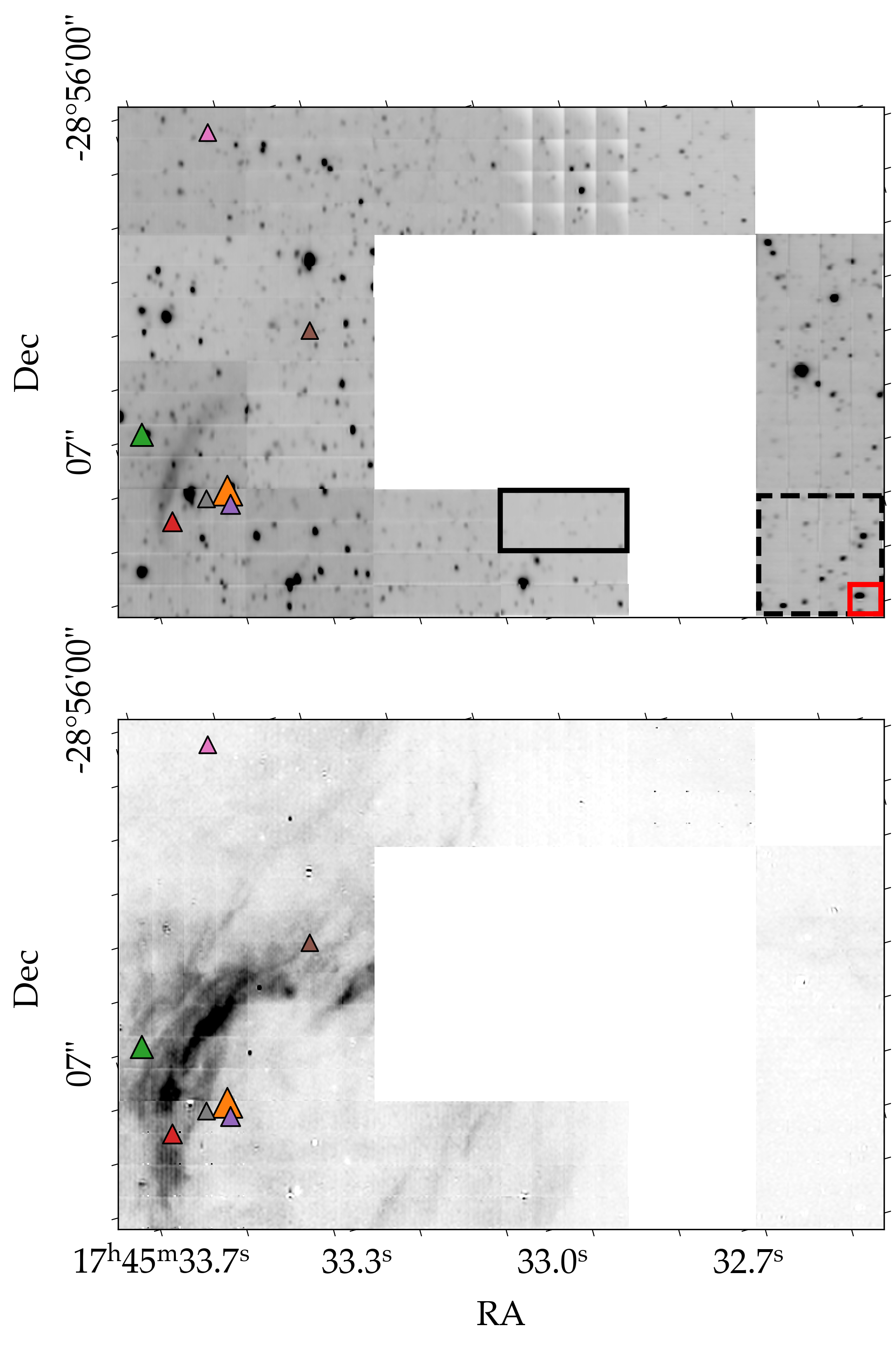}
	\caption{KMOS field of view.  Color points depict identified MYSs with K magnitude brighter than 14.5. The size of each point is proportional to its magnitude. Top: collapsed cube.The red square represents the area covered by a single IFU. The dashed black square represents the area covered by a single IFU after completing a an observational sequence in the MOSAIC mode (16 pointings). The black solid box indicates the size of the fields into which we divided the image to be able to align with the GC catalogs.  Bottom: Br$\gamma$ map obtained after continuum subtraction.}
	\label{fig:brmap}
\end{figure}

KMOS \citep{Kmos_instrument} is equipped with 24 integral field units (IFUs), each offering a field of view of 2.8 by 2.8 arcsec (red square in Fig. \ref{fig:brmap}). The observations where conducted in MOSAIC mode, where the individual IFUs of the 24 arms, are arranged in such a way that with 16 successive telescope pointings, a contiguous rectangular area can be covered (dashed line in Fig.\ref{fig:brmap}). With all 24 IFUs and 16 pointings, it is possible to map an area of $\sim$ 0.8 arcmin$^{2}$. In order to eliminate the infrared emission from the atmosphere, we need to acquire images of an area of the sky relatively free of stars. Typically, these images are obtained by jittering the science observations. However, due to the crowded environment of the GC, the sky observations cannot be obtained from the jittered observations of the target field. Thus, we use a separate field centered on a dark cloud in the GC as the sky (Tab. \ref{tab:obs_data}). \\

\begin{table}
	\tiny
\caption{KMOS observation parametres.}
\label{tab:obs_data}
\begin{tabular}{cccccc}
	\toprule
	Target & RA, Dec (º) & Date & DIT(s) & [$\lambda$($\mu m$)] \\
	\midrule
	Sci & 266.3817, -28.9448 & \multirow{2}{*}{3,6/06/21}  &\multirow{2}{*}{\shortstack{115 \\ NDIT = 1} } &\multirow{2}{*}{ [1.93, 2.50]}\\
	Sky &  266.1914, -28.8962 &  &  &  \\
	\bottomrule
\end{tabular}
\end{table}
We reduced the data using the EsoReflex KMOS pipeline \citep{EsoRex}. The observations were conducted in the same area during two different epochs, with a three-day gap between them. Eight out of the 24 arms were non-operational during both epochs. This had a significant impact, reducing the mosaic mode's coverage by one-third (blank areas in  Fig. \ref{fig:brmap}). Due to a significant misalignment between the two observation epochs, we processed them independently and later merged them using dedicated Python scripts. This approach improved the SNR by approximately 1.5 times. Finally, we fitted the continuum emission with a fourth-degree polynomial. The bottom panel of Fig. \ref{fig:brmap} displays the continuum-subtracted Br$\gamma$ emission.

\subsection{Astrophotometry data}

To provide precise astrometric information for the objects in the KMOS image, we conducted a cross-match with two different catalogs: the GALACTICNUCLEUS survey (hereafter GNS) \citep{GNSI, GNSII} and the 2D proper motion catalog by \cite{LIBRALA2021} (hereafter L21). The combined dataset from these two catalogs is referred to as GNS-L21.

\subsubsection{GNS}

GNS is a near-infrared survey  (J, H, and Ks bands), covering an area of approximately 6000 pc$^{2}$ in the GC with a 0.2$''$ angular resolution. This high spatial resolution is achieved through the application of holographic imaging techniques \citep{holography}. GNS delivers highly precise PSF photometry for over three million stars in the NSD and inner Galactic bar. In order to eliminate foreground stars belonging to the spiral arms and the bulge,  we applied a color cut H$-$Ks > 1.3 \citep[see Fig. 6 in][]{Paco_spiral_arms} . For more details about the GNS, please refer to \cite{GNSI} and \cite{GNSII}

\subsubsection{L21}

We used L21 to provide the stars observed with KMOS with proper motion values. This catalog resulted from two observation sets covering approximately 205 arcmin$^{2}$ \citep[see Fig. 1][]{LIBRALA2021}. These observations, performed in October 2012 and August 2015, utilized the near-infrared channel of the Wide-Field Camera 3 on the HST. Proper motions were calibrated with reference to \textit{Gaia} DR2 \citep{Gaia2}, yielding absolute proper motion measurements for around 830,000 stars. We trimmed the catalog by discarding stars with proper motion errors exceeding 1 mas/yr. For detailed information on data acquisition, reduction, and analysis, refer to \cite{LIBRALA2021}.\footnote{The proper motions catalogs are available at \href{https://academic.oup.com/mnras/article/500/3/3213/5960177}{https://academic.oup.com/mnras/article/500/3/3213/5960177}}

\subsection{Methods }

The analysis of the data was divided into two distinct stages: first, the extraction of spectra from the objects in the field of view from the KMOS data cubes, and second, the identification of these sources in the GNS-L21 catalog.

In the initial stage, we extracted spectra for approximately 300 stars and classified them into two categories: late-type or MYS. Our primary criterion was the presence of CO bandhead absorption, which is characteristic of AGB stars and late-type G, K, M giants \citep{yso_prove, ms_spec}. To identify MYSs, we selected objects whose spectra did not exhibit any detectable CO absorption.

We extracted the spectra using an aperture of 0.5 arcsec, corresponding to the FWHM of the standard stars used for calibration in the KMOS pipeline. We performed background subtraction from a ring with a radius of 2 times the FWHM  and a width 0.2 arcsec around each object to eliminate any Br$\gamma$ interstellar emission which is pervasive in the GC. We customized the background subtraction to avoid using bad pixels or pixels with strong stellar emission.

To estimate the SNR in the MYSs we selected wavelengths intervals in the spectra without any obvious lines and calculated the SNR as the ratio between the mean and standard deviation of the continuum. To ensure the selection of reliable spectra, where the level of the continuum is clearly distinguishable from the background and where we can determine with certainty that no CO lines were present, we classified as MYSs only those stars that, lacking such lines, have an SNR > 15. This SNR level corresponds, in our case, to stars brighter than Ks$\sim$14.5.

 \begin{figure}
 	\centering
 	\includegraphics[width=1\linewidth]{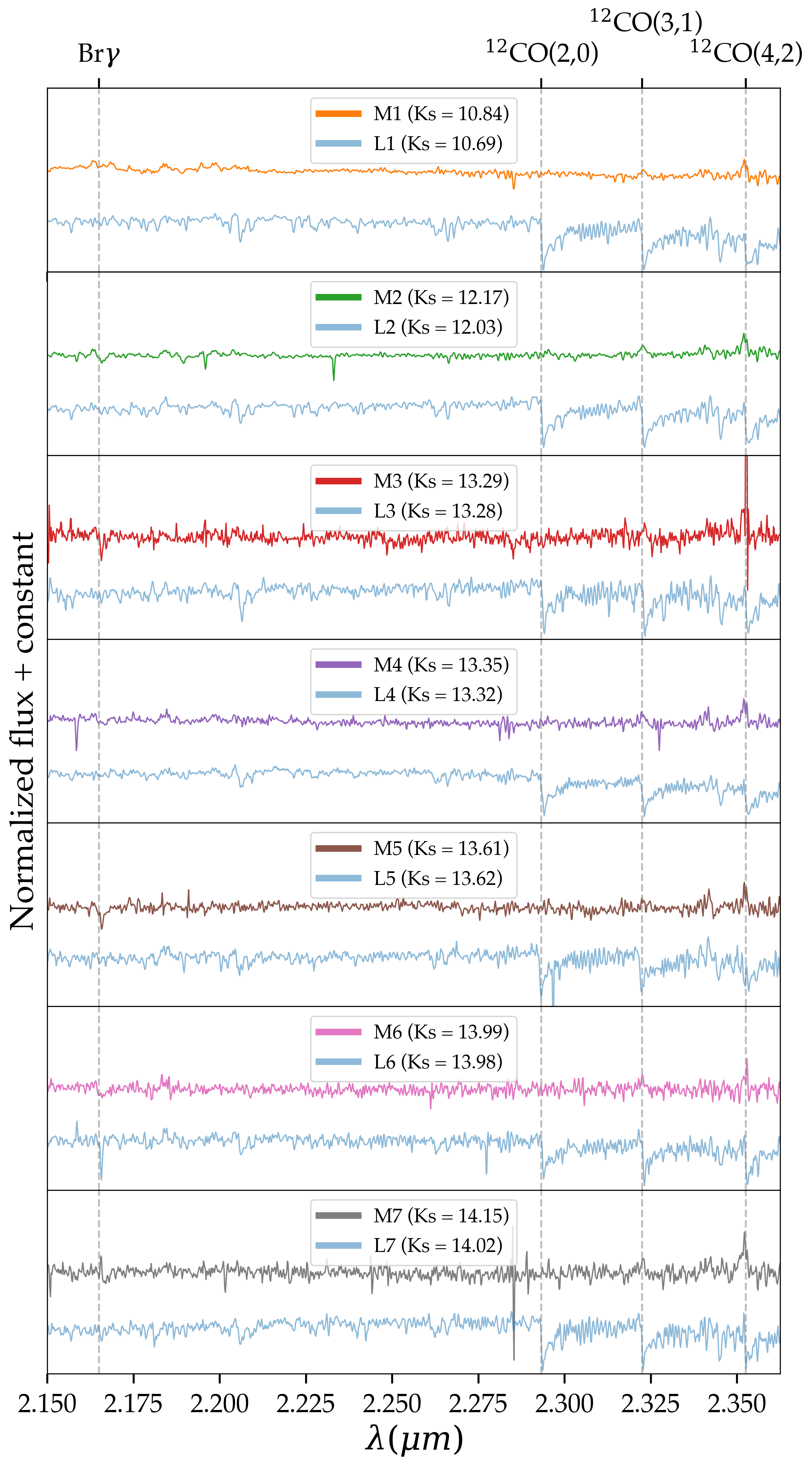}
 	\caption{Spectra of the MYSs. The MYSs are color-coded as in Fig. \ref{fig:brmap} and Fig. \ref{fig:pmcoloryoung}. In blue we show late-type spectra of approximately the same magnitude as that of the MYSs for comparison.}
 	\label{fig:spectrayoung}
 \end{figure}

 In the second part of the analysis, we identified these MYSs in the GNS-L21 catalog, along with the rest of the objects in the field, allowing us to assign proper motions and H and Ks magnitudes. A significant misalignment issue was present among the 16 pointings necessary for a single IFU to complete an entire tile in mosaic mode (indicated by the black dashed square in Fig. \ref{fig:brmap}). Specifically, notable misalignment was observed between the first eight and the last eight pointings across most IFUs. To compensate, we divided the KMOS field into sub-fields, each encompassing eight consecutive pointings (approximately 15 arcsec$^{2}$), where consistent IFU alignment was maintained (black box in Fig. \ref{fig:brmap}, top panel). Using the astroalign Python package \citep{astroalign} and dedicated scripts, we cross-matched objects in these parcels with those in the catalogs. Almost all the  objects identified in the KMOS field of view, including the seven identified MYSs, had counterparts in the GNS-L21 catalog.
 
\begin{figure*}
	\centering
	\includegraphics[width=1\linewidth]{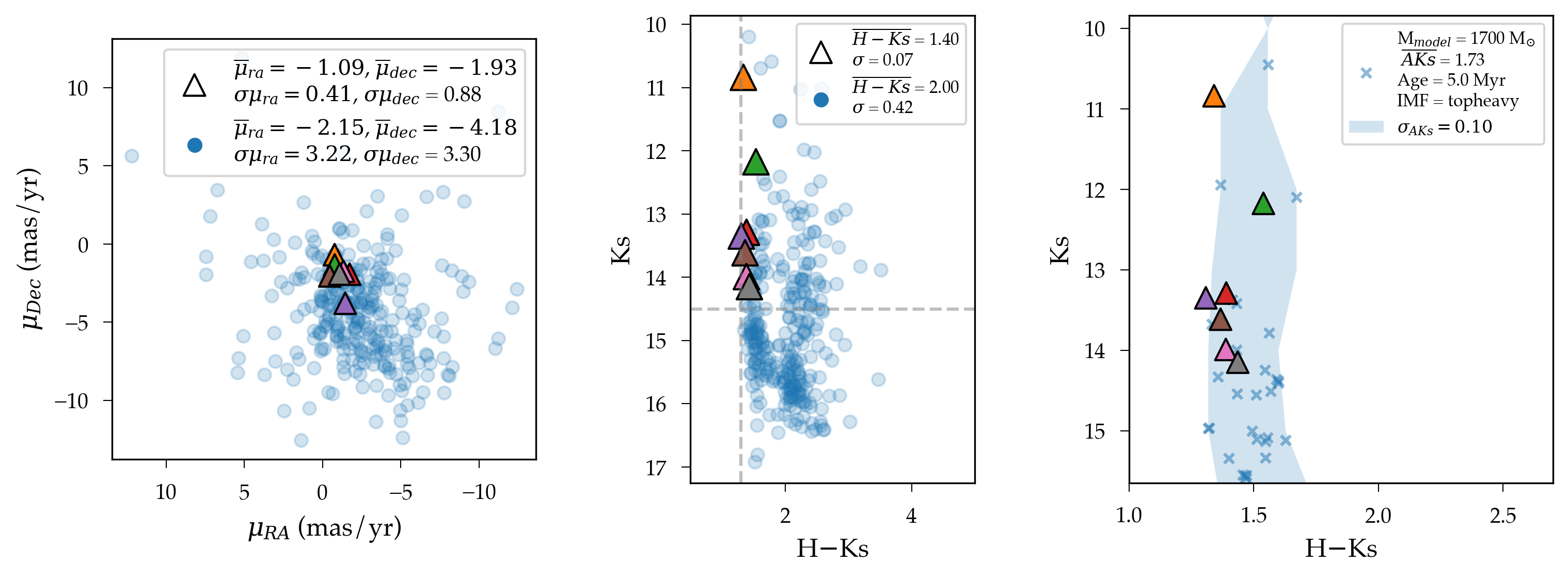}
	\caption{KMOS objects proper motions and CMD, and  cluster simulation. The triangles represent the identified MYS and the blue circles represent the rest of the objects in the field. Left: Vector-point diagram. Middle: CMD. The vertical dashed line marks the color-cut we made to exclude foreground stars (H$-$Ks = 1.3). The horizontal dashed line marks the magnitude cut we made to exclude stars with low spectroscopic SNR (Ks = 14.5). Right: Cluster simulation. The blue crosses represent members of a simulated cluster. The legend box displays the main features of the cluster. The shaded area represents the uncertainty in the position in the CMD for the members of the simulated cluster due to variations in extinction. }
	\label{fig:pmcoloryoung}
\end{figure*}

\begin{table}
	\caption{MYSs parametres. Uncertainties in the propermotions as they appear in L21. Uncertainties in the magnitudes from GNS are $\lesssim$ 0.01 therefore irrelevant for this analysis. }
	\tiny
	\label{tab:ysos}
	\begin{tabular}{cccccc}
		\toprule
		ID & RA, Dec (º) & $\mu_{RA}$(mas/yr) & $\mu_{Dec}$(mas/yr) & H & Ks \\
		\midrule
		M1 & 266.3865, -28.9380& -0.77$\pm$0.53 & -0.69$\pm$0.38 & 12.18 & 10.84 \\
		M2 & 266.3857, -28.9357  & -0.78$\pm$0.46 & -1.35$\pm$0.35 & 13.71 & 12.17 \\
		M3 & 266.3877, -28.9369 & -1.73$\pm$0.51 & -1.94$\pm$0.51 & 14.68 & 13.29 \\
		M4 & 266.3868, -28.9381 & -1.45$\pm$0.15 & -3.79$\pm$0.26 & 14.66 & 13.35 \\
		M5 & 266.3818, -28.9388 & -0.46$\pm$0.44 & -2.05$\pm$0.29 & 14.98 & 13.61 \\
		M6 & 266.3774, -28.9353 & -1.34$\pm$0.22 & -1.74$\pm$0.28 & 15.38 & 13.99 \\
		M7 &266.3869, -28.9375 & -1.09$\pm$0.28 & -1.93$\pm$0.25 & 15.58 & 14.15 \\
		\bottomrule
	\end{tabular}
\end{table}

\section{Results}

We display the identified MYSs in the KMOS image with counterparts in the GNS-L21 catalog in the left and central panels of Fig. \ref{fig:pmcoloryoung}. The triangles represent the identified MYSs, color-coded as in Fig. \ref{fig:brmap} and Fig. \ref{fig:spectrayoung}. The coordinates, proper motions, and magnitudes of the MYSs are shown in Table \ref{tab:ysos}.

In the left panel of Fig. \ref{fig:pmcoloryoung}, we show a vector-point diagram of their equatorial proper motions, while the middle panel shows a CMD. Notably, in Fig. \ref{fig:pmcoloryoung} left panel, the velocity dispersion in the MYSs group is significantly smaller than the one corresponding to the rest of the stars in the field. In the middle panel, we observe that all the MYSs display a similar color, indicating that they are affected by similar extinction. This supports the idea that these stars are located at a similar depth in the NSD. The similar values in velocity, position in the sky, line-of-sight distance, and the fact that all seven members  are a massive, young stars strongly indicate that this group shares a common origin. Moreover, this group is situated within a recognized HII region \citep{HII_regions}, displaying significant Br$\gamma$ emission, as illustrated in the lower panel of Fig. \ref{fig:brmap}, providing additional evidence for recent, in situ star formation \citep{Hankins:2019sw}.

If this co-moving group is part of a larger cluster or stellar association, we can estimate its mass by comparing it with simulated clusters. The Python package Spisea \citep{Spisea} allows the generation of single-age, single-metallicity clusters, and we utilized it to create models for comparison with the selected co-moving group. To generate a cluster model, we needed to set several parameters as input for Spisea: distance, extinction, initial mass function (IMF), age, and metallicity. We adopted a distance of 8.25 kpc \citep{GC_distance_gravity}. We obtained the extinction value for each star in the MYS group from the catalog provided by \cite{GNSIV}. Estimating a single value for the remaining variables was challenging with the present data set, so we ran a series of simulations with various combinations. For the ages, we selected four different values ranging from 2.5 Myr, which is the age of the youngest cluster in the NSD, the Arches \citep{Arches_2.5Myr}, to 10 Myr, which is the maximum estimated dissolving time for a massive cluster in the GC \citep{cluster_dissolution}. For the metalicity, we considered [M/H] = 0 and [M/H] = 0.3 \citep{Arches_Metal, paco_nature_2020, NSD_metal, NSD_trans_metal}. Additionally, we used two different IMFs: the broken power-law derived by \cite{IMF} and the top-heavy one derived by \cite{Arches_IMF}. This resulted in a total of 16 possible configurations, each run 100 times, producing a total of 1600 simulations.

For each of these simulations, we defined a magnitude interval to compare the model with the co-moving group of MYSs, using the brighter and fainter magnitudes of the MYSs group. Then, we randomly assigned either a very low or a very high initial mass to the model and compared the number of stars in the magnitude interval with the number of stars in the MYS group. If the number of stars in this interval differed from the number of stars in the MYS group, we adjusted the mass of the cluster, either increasing it or decreasing it. This process was repeated, gradually adjusting the mass with $\pm$1\% increments until the number of stars in the magnitude interval for the model equaled the number of stars in the MYS group. Due to completeness limitations, this method allows us to estimate only a lower limit for the parent cluster of the co-moving group, yielding a value of M$_{estimated}$\,=\,1744\,$\pm$\,723\,M$_{\odot}$. In Fig. \ref{fig:pmcoloryoung} right panel, we present the CMD of one of these 1600 simulated models alongside the values of the real MYSs group for comparison.

\begin{figure}
	\centering
	\includegraphics[width=0.7\linewidth]{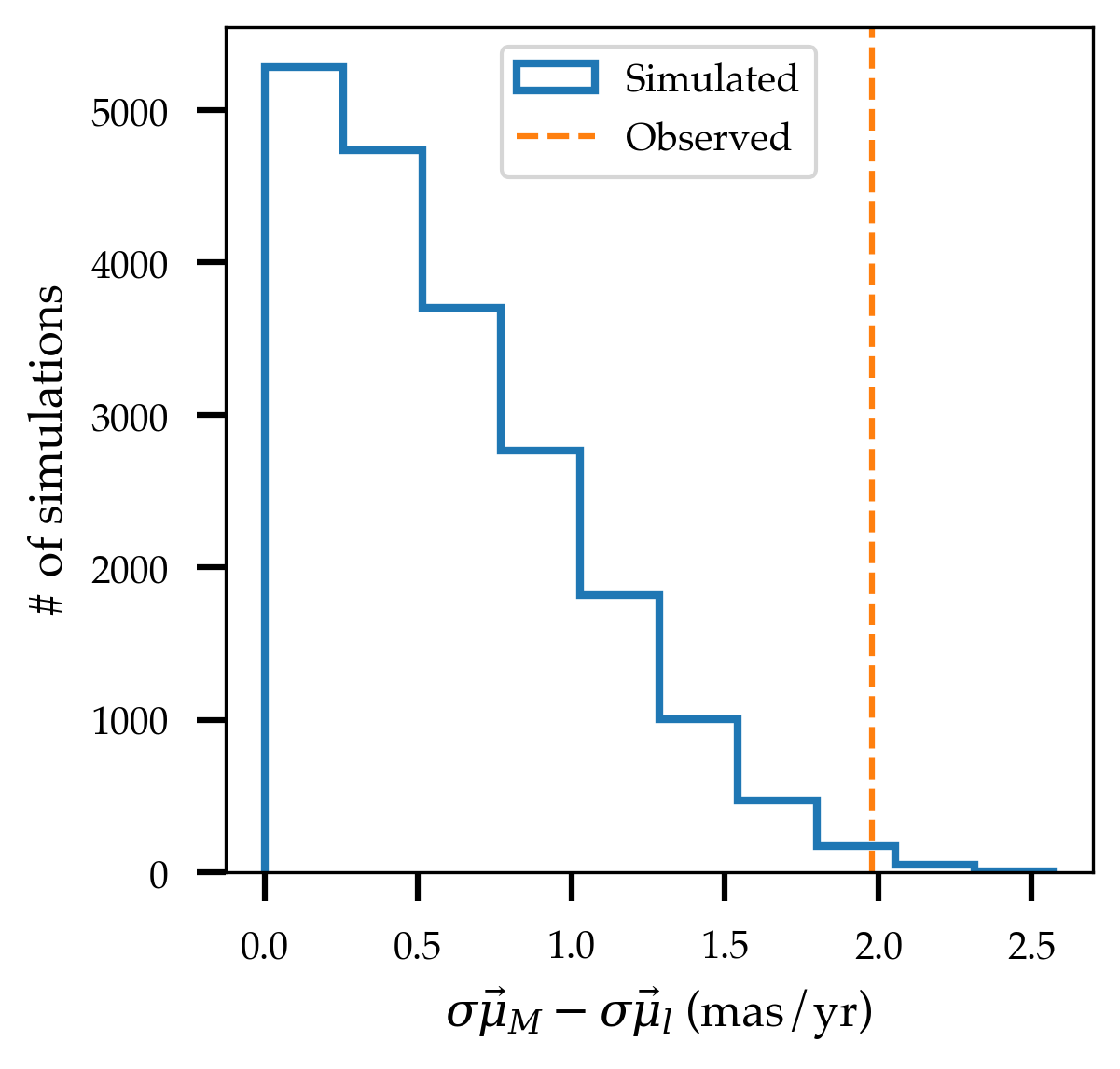}
	\caption{Velocity dispersion simulations. The blue histogram represent the difference in velocity dispersion between simulated populations of MYSs and late type stars, that where created by shuffling the velocities among the observed sample of stars. The orange dashed line represents the value between the observed MYS group and the late-type stars.}
	\label{fig:simsig}
\end{figure}

To investigate the probability of finding a group of seven stars classified as MYSs with a velocity dispersion as small as the one of the group found here formed purely by a random association of stars in the field, we conducted a permutation test. The used statistical test was the difference in the values of the velocity dispersion between the MYSs group and the rest of the field stars classified as late type, considering only stars with Ks magnitudes brighter than Ks = 14.5. The observed difference in the velocity dispersion for the real data is |$\sigma\overrightarrow{\mu}{yso} -\sigma\overrightarrow{\mu}{late}$| = 1.98 mas/yr (dashed line in Fig.\ref{fig:simsig}). We then randomly shuffled the velocities between all the  stars and compared the differences in the velocity dispersions. This process was repeated 20,000 times, and the results are shown in Fig. \ref{fig:simsig}. Only about 0.4\% of the 20,000 simulated MYS groups resulted with a |$\sigma\overrightarrow{\mu}{yso} -\sigma\overrightarrow{\mu}{late}$| equal or larger than the real data (blue histogram to the right of the dashed line in Fig. \ref{fig:simsig}), indicating that the observed association of MYSs is highly unlikely to have occurred by chance.

\begin{figure}
	\centering
	\includegraphics[width=1\linewidth]{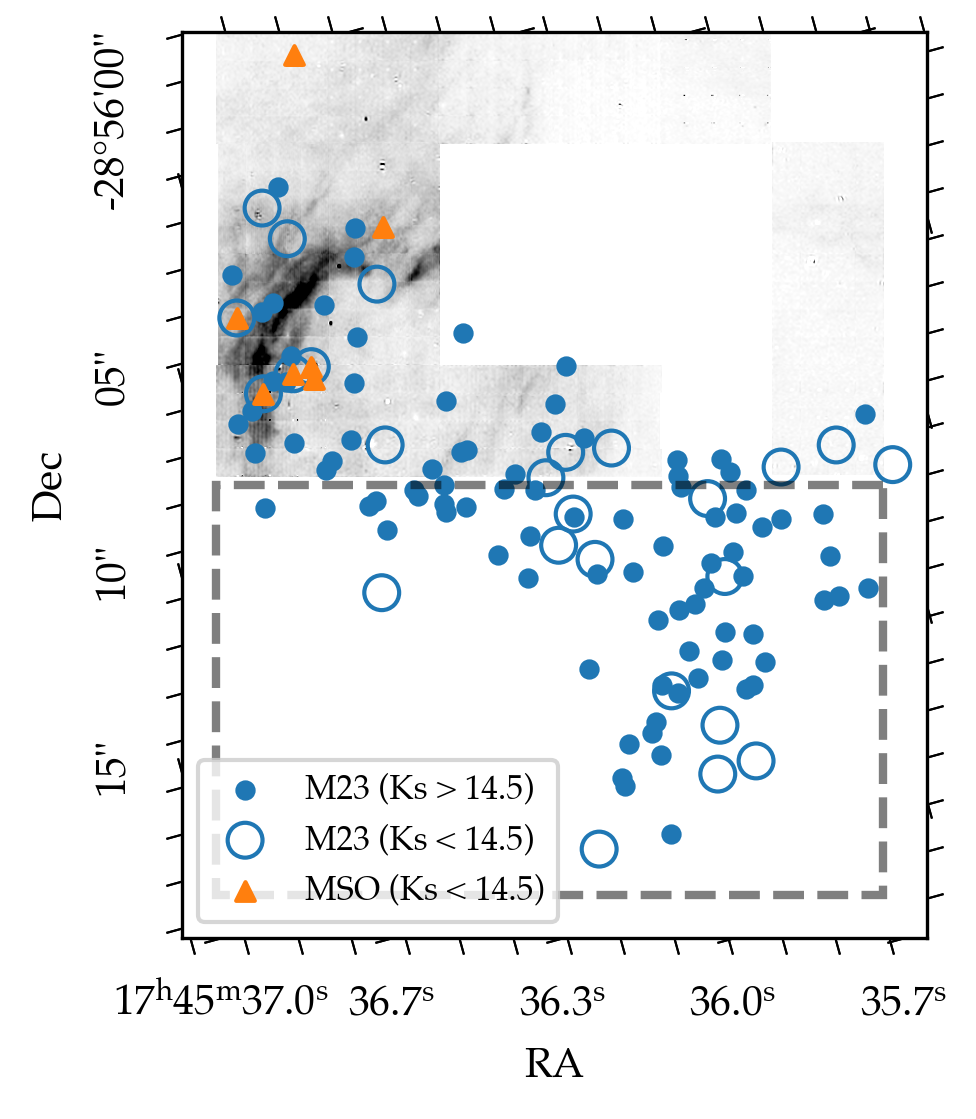}
	\caption{Co-moving group and MYSs on a Br$\gamma$ emission map. Orange triangle represent the MYSs found in the KMOS field of view. Blue points and circles represent the co-moving group presented in \cite{comoving_arxiv}. The dashed square outlines the region scheduled for observations using the KMOS instrument during the ESO period P113.}
	\label{fig:clusterbrg}
\end{figure}

It is worth mentioning that recently, another co-moving group has been identified by \cite{comoving_arxiv} (Fig. \ref{fig:clusterbrg}), and it partially overlaps with the one reported by \cite{Ban_cluster}. Notably, these two co-moving groups were identified using different catalogs and methodologies. Both co-moving groups share common elements with the co-moving group of MYSs presented here.

\section{Conclusions}

\cite{Ban_cluster} reported on a group of seven co-moving stars potentially associated with the Galactic Center H1 HII region, that is located at a projected distance of only about 11\,pc from Sagittarius\,A* \citep[see, e.g.,][]{HII_regions,Hankins:2019sw},  and with a strong Paschen-$\alpha$ emitting star, the O super-/hyperrgiant P114  \citep{Blue_SG,Massive_stars}. Subsequently \cite{comoving_arxiv} exploited the HST WFC3 proper motion catalogue of \cite{LIBRALA2021} and found a group of 120 co-moving stars associated with P114 \citep[ID14966 in][]{LIBRALA2021} that may indicate the presence of a young cluster of a few thousand solar masses. Most of the star in the co-moving group identified in \cite{Ban_cluster} were part of the one reported by \cite{comoving_arxiv}. Notably, these two co-moving groups were identified using different catalogs and methodologies.


Here we present spectroscopic observations of an area associated with this potential cluster or stellar association. We assigned proper motions and magnitudes to approximately 300 stars, of which about one-third had sufficient SNR to classify them as late-type giants or early type massive young stars. We have found six additional young massive stars in addition to P114. Five of those stars lie within a radius of  $\sim 5"$ of projected distance (about 0.20\,pc at the distance of the GC) of P114, with the sixth one at  $\sim15"$ ($0.60$\,pc) (Fig.\,\ref{fig:brmap}). The massive young stars lie very close to each other in the proper motion diagram (left panel in Fig.\,\ref{fig:pmcoloryoung}). Furthermore, all seven massive stars have the same $HKs$ colors within the uncertainties, which supports the idea  that they also lie close to each other along the line-of-sight  (central panel in Fig.\,\ref{fig:pmcoloryoung}). It is worth noting that this co-moving group is situated very close to the more probable orbit for the Arches cluster as derived by \cite{Arches_Hosek}. \citep[see also Fig. 1 in][]{comoving_arxiv}. Additionally, the projected distance from the Arches, approximately 20 pc, aligns well with the tidal tail simulation presented in \cite{Arches_Quint_tails}. Considering these coincidences, the possibility that this group of MYSs may represent traces of a tidal tail associated with the Arches cluster can not be ruled out. Deeper and more comprehensive observations of the area are necessary to over the unobserved areas around the main group and acquire spectra with higher SNR, enabling us to classify the ages of fainter sources.

Given these results and the co-existence with an HII region in the plane of the sky, we believe that the evidence is now sufficiently strong to speak of a newly discovered young cluster at the GC.  While we have known for several decades only of three young clusters in the GC (Arches, Quintuplet, central parsec) plus a few hundred known massive stars distributed through the field \citep{Massive_stars}, proper motion studies of the GC are now sufficiently precise and complete that we can pinpoint potential clusters via kinematics and then confirm them spectroscopically. This opens a new window onto studying star formation in this unique astrophysical environment. Only a $\lesssim$\,5\% of the area of the nuclear stellar disk has been exploited with this method, with only one candidate cluster confirmed spectroscopically so far. This suggests that we may still be able to pinpoint a dozen or more new clusters. All of of them will probably be younger than $\sim$10\,Myr, because significantly older clusters are expected to have dissolved \citep{cluster_dissolution}. After having found these new clusters, we will be able to study them in more detail. The perhaps most important question to address will be: Is the IMF at the GC indeed different from the Galactic disc, as indicated by the observations of the three known clusters and also motivated by theoretical considerations \citep[e.g.][]{Morris:1993ve,Bartko:2010fk,Husmann:2012vn,Arches_IMF}?

\bibliographystyle{aa} 
\bibliography{Kmos_cit}
\end{document}